\documentclass[twocolumn,showpacs,preprintnumbers,amsmath,amssymb,floatfix]{revtex4}

\usepackage{graphicx}
\usepackage{dcolumn}
\usepackage{bm}
\usepackage{hyperref}
\usepackage{colordvi}

\begin{document}

\hsize\textwidth\columnwidth\hsize
\csname@twocolumnfalse\endcsname

\title{Three-dimensional imaging of colloidal glasses under steady shear}

\author{R.~Besseling$^1$, Eric R.~Weeks$^2$, A.~B.~Schofield$^1$, W.~C.~K.~Poon$^1$}

\address{$^1$Scottish Universities Physics Alliance (SUPA) and School of Physics, \\ The University
of Edinburgh, Kings Buildings, Mayfield Road, Edinburgh EH9 3JZ, United Kingdom.\\ $^2$ Physics Department, Emory
University, Atlanta, Georgia 30322, USA.}

\date{\today}

\begin{abstract}
Using fast confocal microscopy we image the three-dimensional dynamics of particles in a yielded hard-sphere
colloidal glass under steady shear. The structural relaxation, observed in regions with uniform shear, is nearly
isotropic but is distinctly different from that of quiescent metastable colloidal fluids. The inverse relaxation
time $\tau_\alpha^{-1}$ and diffusion constant $D$, as functions of the {\it local} shear rate $\dot{\gamma}$,
show marked shear thinning with $\tau_\alpha^{-1} \propto D \propto \dot{\gamma}^{0.8}$ over more than two decades
in $\dot{\gamma}$. In contrast, the {\it global} rheology of the system displays Herschel-Bulkley behavior. We
discuss the possible role of large scale shear localization and other mechanisms in generating this difference.

\end{abstract}


\maketitle

\narrowtext \noindent

Glassy materials are ubiquitous in nature and in industry; examples range from molecular and metallic glasses
\cite{molglass,metglass} to soft glasses like colloidal suspensions, emulsions and foams
\cite{rheoLarsonBook,SollichCatesPRL97_softglassyrheology}. Of special importance, both fundamentally and
practically, is their rheological behavior. Glasses have liquid-like microstructure, but solid-like mechanical
behavior. At low applied stress, they have finite shear moduli, but at sufficient stresses they yield and display
highly nonlinear flow behavior.

Among the many open issues in {\it nonlinear} glassy rheology, steady shear stands as the simplest example, yet it
is far from being fully understood. Theories
\cite{SollichCatesPRL97_softglassyrheology,Berthier,FuchsCatesPRL02_nonlinearheo,Falk} have invoked various
mechanisms for shear-induced relaxation of initially arrested structures, predicting a variety of constitutive
relations. Significantly, all these theories assume globally uniform shear. Simulations, so far the main tool to
check directly the relation between microscopic behavior and macroscopic flow, reveal spatially heterogeneous
relaxation \cite{Falk,YamamotoOnukiPRE98PRL98} and strong shear thinning \cite{BerthierJChemPhys02,Miyazaki04}.
Experiments are just starting to address microscopic dynamics under shear, but have been limited to coarse-grained
data, two dimensional (2D) or interrupted flows, or ordering phenomena
\cite{Petekidis,Hebraud,rheoimage,BonnPRL02_laponiteagingrejuvenationrheo}. Moreover, experiments imaging {\it
global} flow \cite{CoussotBecubanding} as well as boundary driven simulations of Lennard-Jones (LJ) glasses
\cite{VarnikBarratPRL03_LJshearlocali} show that (soft) glasses often exhibit shear localization, which can not be
described by simple constitutive laws.

In this Letter we report a three-dimensional (3D) imaging study of the microscopic relaxation in a colloidal glass
under steady shear. The relaxation is nearly isotropic but different from that of unsheared colloidal fluids. The
inverse relaxation time $\tau_{\alpha}^{-1}$ and the diffusion constant $D$ show marked shear thinning as a
function of the {\it local} shear rate $\dot{\gamma}$: $\tau_\alpha^{-1} \propto D \propto \dot{\gamma}^{0.8}$. We
find that this local behavior contrasts significantly with the {\it global} rheology, which shows Herschel-Bulkley
behavior.

We used sterically-stabilized polymethylmethacrylate (PMMA) particles (radius $a=850$nm, measured by light
scattering, polydispersity $\lesssim 10\%$ \cite{fn_polydisp}) fluorescently labelled with nitrobenzoxadiazole and
suspended in a mixture of cycloheptyl bromide and decalin (viscosity 2.6 mPa$\cdot$s) for density and refractive
index matching. In this medium particles acquire a small charge \cite{YethirajNature03_CHBmodelsystem} which is
largely screened by adding 4~mM tetrabutylammonium chloride, giving nearly hard-sphere (HS) behavior, with a glass
transition at volume fraction $\phi_g \simeq 0.58$ (determined from mean-squared displacements)
\cite{fn_expversusHS}; we work at $\phi \simeq 0.62$ (measured by imaging). The reduced shear rate, or P\'{e}clet
number, is $\mbox{Pe}=4a^2\dot{\gamma}/D_0= 24 \dot{\gamma}\tau_B$, with $D_0$ the bare diffusion coefficient and
$\tau_B=a^2/6 D_0$= 1.24~s the Brownian time in our system. Our experiments cover the range $0.005 \lesssim
\mbox{Pe} \lesssim 1 $.

We employ a linear parallel-plate shear cell with a plate separation $\sim 400-800$ $\mu$m, parallel to $\pm 5 \mu
m$ over a $\sim 200$~mm$^2$ drop of colloid confined between the plates by surface tension. We define $x$, $y$ and
$z$ as the velocity, vorticity (or neutral) and gradient directions respectively. The top plate is driven at
$0.05-10$ ~$\mu$m/s by a mechanical actuator with magnetic encoder, and steady shear is applied up to a total
accumulated strain of $\Delta \gamma \simeq 1000 \%$. Wall slip and wall-induced ordering were prevented by
coating the slides with 1-3 disordered layers of particles. A solvent bath minimized evaporation.

A $30\times30\times15$~$\mu$m$^3$ volume in the drop (containing $N \sim 3000$ particles) was imaged from below as
a stack of $75$ slices using a fast confocal scanner (VT-Eye, Visitech International) and a Nikkon TE Eclipse
$300$ inverted microscope. The scanning of each 3D stack took $1.7$~s. Particles were located with resolution
$\delta x,\delta y \sim 30$~nm and $\delta z\sim 90$~nm \cite{CrockerGrierJColIntSc96_tracking}. Tracking from
frame to frame was achieved by first subtracting from the raw coordinates a time ($t$) dependent $x$-displacement
profile $\Delta x(z,t)$, measured via correlation analysis of raw images, and adding this back after particle
tracking. The resulting $x$-displacements over a given time interval $dt$, $\{\Delta x_{i}(z_i,dt)\}$ ($i = 1$ to
$N$), always have an average linear dependence on $z$. From this we checked that the sample in our imaged volume
was indeed subjected to uniform shear, and measured the actual (local) shear rate $\dot{\gamma}$, which may differ
from the applied (global) rate $\dot{\gamma}_a$ due to shear localization and the presence of jammed regions. We
will return to this point; for now we focus on steady states with a linear velocity profile in a region from $15
-30$~$\mu$m above the cover slide. When present, strong decay in the shear rate occurs at least $\Delta z \sim
20a$ away from imaged regions. We also checked, via bond-order analysis \cite{fn_localorder}, that shear-induced
crystallization was absent for our range of $\dot{\gamma}$ \cite{HSshearorder}.

\begin{figure}

\scalebox{0.45}{\includegraphics{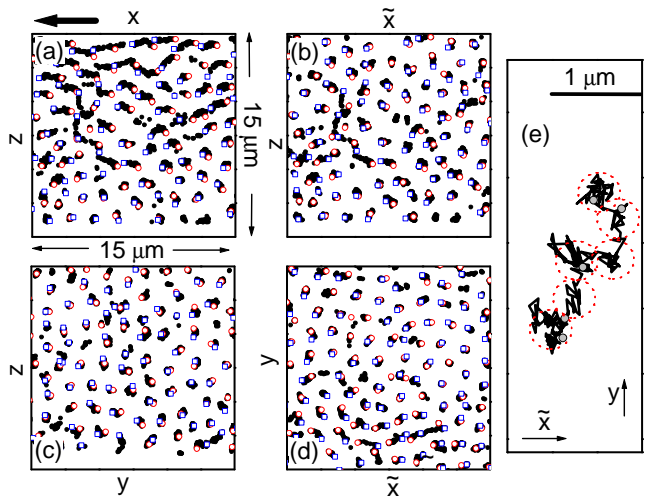}}

\caption{Colloid trajectories for $\dot{\gamma} = 9.3 \times 10^{-4}$~s$^{-1}$. (a) 1.5~$\mu$m thick slice in the
$x, z$ plane for 160 s; The start of each trajectory is shown by \Red{$\circ$}, the end by \Blue{$\square$}. The
big arrow marks the shear direction. (b) As in (a) but in the de-sheared, $\tilde{x}, z$, reference frame, with
$\tilde{x}_i=x_i-\dot{\gamma}\int_0^t z_i(t') dt'$. (c) $y, z$ plane over 160~s. (d) $\tilde{x}, y$ plane over
160~s. (e) Single trajectory in the $\tilde{x}, y$ plane over 800~s. Dotted circles mark rattling in several cages
({\it not} the particle size), grey dots show the locations at $t=0,200,400,600,800$~s.}
      \label{fig_tra}
\end{figure}

Figure \ref{fig_tra}(a) shows the trajectories in an $x, z$ slice at $\dot{\gamma}=0.93 \times 10^{-3}$~s$^{-1}$.
The displacement gradient due to shear is evident. To highlight the shear-induced dynamics, we show in
Fig.~\ref{fig_tra}(b) the {\it non-affine} component of the motion obtained by subtracting the uniform shear via
$\tilde{x}_i=x_i-\dot{\gamma}\int_0^t z_i(t') dt'$. Considerable shear-induced non-affine displacements are seen
in this plane as well as in the other planes, Figs. \ref{fig_tra}(c,d). On the time scale considered here, these
rearrangements are heterogeneous, somewhat similar to observations in {\it quiescent} concentrated colloidal
fluids for $\phi<\phi_g$ \cite{Weeks_Kegel}. Zooming in on a single particle, Fig.~\ref{fig_tra}(e), we observe
that its dynamics under shear consists of intervals of cage `rattling', interrupted by shear-induced plastic
cage-breaking events.

\begin{figure}
\scalebox{0.45}{\includegraphics{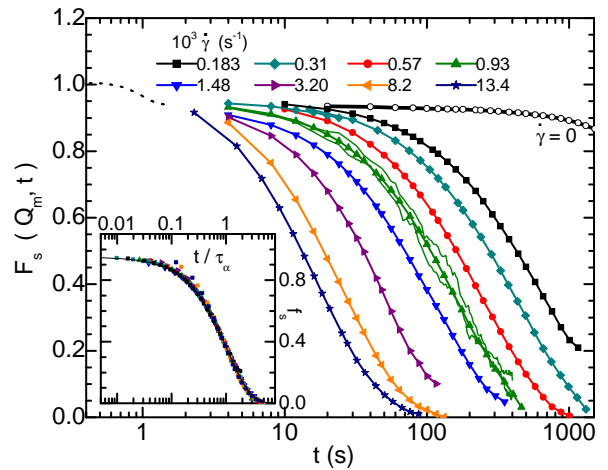}}

\caption{Incoherent scattering functions $F_s(Q_m,t)$, with $\dot{\gamma}$ increasing from right to left. Lines
for $\dot{\gamma}=0.93 \times 10^{-3}$~s$^{-1}$ show two curves used in the average with start times $t_0$ spaced
by $180$s. The dashed line schematizes initial relaxation. Inset: data collapse using $f_s(Q_m,t/\tau_{\alpha})$.
Line: $f_s \propto \exp(-t/\tau_\alpha)$.} \label{figFself}
\end{figure}

Next, we study the relaxation via the incoherent scattering function, $F_s(Q,t)=\langle
\cos(Q[y_i(t_0+t)-y_i(t_0)]) \rangle_{i,t_0}$, at a scattering vector $Q =Q_m \simeq 3.8a^{-1}$ where the data's
structure factor $S(Q)$ shows a peak. In Fig.~\ref{figFself} we show selected results for $\vec{Q}\parallel y$,
but the results (not shown) for $\vec{Q}\parallel z$ and $x$, using the non-affine displacements $\tilde{x}_i$ for
the latter, are similar. $F_s$ for the quiescent glass ($\dot{\gamma} = 0$) hardly decays over our observation
window, reflecting the caging of particles by their neighbors; at longer times we observed aging, as in other
studies \cite{BonnPRL02_laponiteagingrejuvenationrheo,CourtlandWeeksJPCM03_agingglassconfocal}. The short time
decay due to initial cage exploration ($t \lesssim \tau_B$ \cite{MegenPRE98_tracersinglass}, dashed line in
Fig.~\ref{figFself}) is inaccessible to us. At small $\dot{\gamma}$, $F_s$ at short times still exhibits a
plateau, in agreement with the caging in Fig.~\ref{fig_tra}(e). As $\dot{\gamma}$ increases, this plateau shrinks
and for the highest $\dot{\gamma}$ it vanishes and likely merges with the short time decay. At longer times, $F_s$
decays strongly for all $\dot{\gamma} \neq 0$, marking shear-induced structural relaxation and cage
rearrangements. The structural ($\alpha$-)relaxation time $\tau_\alpha$, defined by
$F_s(Q_m,t=\tau_{\alpha})=e^{-1}$, decreases on increasing $\dot{\gamma}$. Importantly, $F_s$ is independent of
the starting time $t_0$ (see data for $\dot{\gamma}=0.93 \times 10^{-3}$~s$^{-1}$), i.e., a stationary state is
achieved.

Our data confirms the theoretically-predicted `time-shear superposition principle'
\cite{Berthier,FuchsCatesPRL02_nonlinearheo}: when time is scaled by $\tau_\alpha$, the $\alpha$-relaxation
follows a master curve $f_s(Q,t/\tau_\alpha)$, Fig.~\ref{figFself} inset. As in LJ simulations
\cite{BerthierJChemPhys02}, our $f_s$ is a pure exponential. This differentiates a shear-melted glass from a dense
HS colloidal fluid at $\phi<\phi_g$ and $\dot{\gamma}=0$, where we find stretched exponential behavior for $F_s(Q
\gtrsim Q_m/2)$, as can also be deduced from \cite{MegenPRE98_tracersinglass}.

\begin{figure}
\scalebox{0.4}{\includegraphics{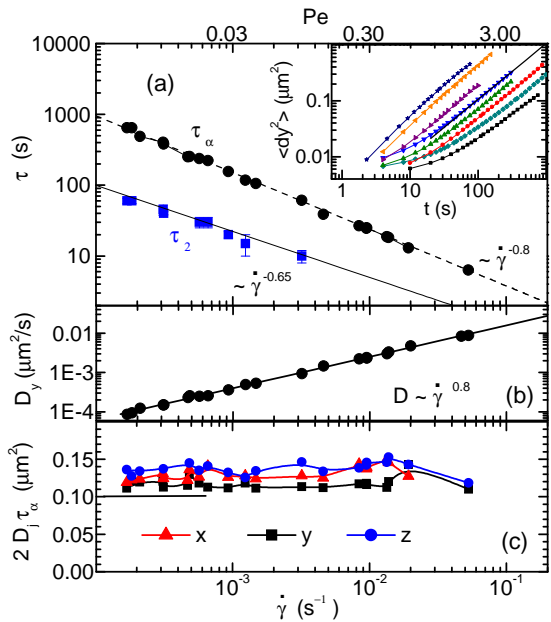}} \caption{(a) Structural relaxation time ($\bullet$) and the
characteristic time $\tau_2$ for the crossover from caged to diffusive behavior (\Blue{$\blacksquare$}) vs.
$\dot{\gamma}$; dashed line: $\tau_{\alpha} \propto \dot{\gamma}^{-0.8}$; full line: $\tau_2 \propto
\dot{\gamma}^{-0.65}$. Inset: mean square displacement in the vorticity direction for shear rates as in
Fig.~\ref{figFself}. Line: $\langle dy^2(t) \rangle = 2 D_y t$ for $\dot{\gamma}=1.48 \times 10^{-3}$~s$^{-1}$.
(b) ($\bullet$) Diffusion constant $D_y$ vs. $\dot{\gamma}$. Line: $D_y \propto \dot{\gamma}^{0.8}$. (c) The
scaled diffusion constant $2 D_j \tau_\alpha \simeq \langle dr_j^2(\tau_\alpha) \rangle $ vs. $\dot{\gamma}$ for
$j=x,y,z$. Line: the value $\langle dy^2(\tau_\alpha) \rangle =2/Q_m^2$ expected for gaussian behavior.}
\label{figversusgammadot}
\end{figure}

Figure \ref{figversusgammadot}(a) shows the dependence of $\tau_\alpha$ on the shear rate. It exhibits a power law
$\tau_{\alpha} \propto \dot{\gamma}^{-\nu}$ with $\nu=0.80 \pm 0.01$ \cite{fn_tauaphi}, independent of the
criterion or $Q$ used to determine $\tau_a$. This behavior means that the accumulated strain at $\tau_\alpha$ is
not constant but varies as $\dot{\gamma} \tau_\alpha \propto \dot{\gamma}^{0.2}$. The data are consistent with a
schematic model \cite{Berthier} for driven glasses and also match the `creep' behavior of a driven particle in a
correlated random potential \cite{Horner96}. We note that an 'entropic barrier hopping' model
\cite{KobelevSchweizer_shear}, without any 'ideal' glass divergences, shows a very similar dependence of the
'hopping' time on $\dot{\gamma}$. Below we discuss the rheological implications of this behavior.

Turning to the mean squared displacement (MSD) $\langle dy^2(t) \rangle$, Fig.~\ref{figversusgammadot}(a) inset,
we see that it exhibits a crossover from caged to diffusive motion for $\sqrt{\langle dy^2 \rangle} /a \simeq
0.15$ ($\langle dy^2 \rangle \simeq 0.017$ $\mu$m$^2$), in reasonable agreement with the `Lindemann parameter'
measuring the cage rattling at the quiescent glass transition \cite{fn_expversusHS}. The long time diffusion
constant $D_y$, Fig.~\ref{figversusgammadot}(b), follows the relaxation rate $D_y \propto \tau_\alpha^{-1} \propto
\dot{\gamma}^{0.8}$, and {\it not} the shear rate $\dot{\gamma}$. To show this more clearly and also address the
anisotropy in the dynamics, we plot in Fig.~\ref{figversusgammadot}(c) the product $2D_j \tau_{\alpha}$ for the
three directions ($j=x,y,z$) along with the value $\langle dy^2(\tau_{\alpha}) \rangle=2/Q_m^2$ expected from a
gaussian approximation $F_s(Q_m,t) \simeq e^{-Q_m^2 \langle dy^2(t) \rangle /2}$ \cite{MegenPRE98_tracersinglass}.
The value for $2D_y\tau_{\alpha}$ agrees well with $2Q_m^{-2}$ and this gaussian long time behavior also occurs in
the other directions \cite{fn_tauj}. We again stress the difference with quiescent fluids at $\phi<\phi_g$, which
always show $D \tau_{\alpha}< Q^{-2}$ for $Q \gtrsim Q_m$. Figure ~\ref{figversusgammadot}(c) also shows that the
diffusion constants exhibit only a mild anisotropy: while $D_z > D_{x,y}$, the difference is $\lesssim 20\%$.
Similar or even smaller anisotropy has been observed in simulations of sheared, glassy systems
\cite{YamamotoOnukiPRE98PRL98,Miyazaki04} and colloidal fluids \cite{FossBradyJFM99_sheardiffusion}. Isotropic
shear-induced diffusion is also seen in {\it dilute} suspensions \cite{QiuChaikinPRL88_diffusioninshear}. However,
sheared {\it non-Brownian} suspensions ($\mbox{Pe} \rightarrow \infty$) show a marked anisotropy
($D^{\infty}_x/D^{\infty}_{y,z} \sim 8$) \cite{BreedveldJCP02_nonbrowndiffusion}, with $D^{\infty} \propto
\dot{\gamma}$.

\begin{figure}[h]

\scalebox{0.45}{\includegraphics{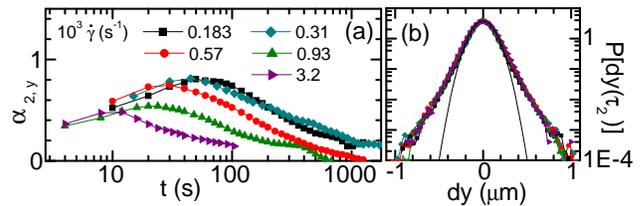}} \caption{(a) Nongaussian parameter $\alpha_2(t)$ of the probability
distribution $P[dy(t)]$, for several $\dot{\gamma}$. (b) $P[dy(t=\tau_2(\dot{\gamma}))]$ for the corresponding
$\dot{\gamma}$, showing a near collapse of the data (each involving $>10^5$ displacements). Line: best gaussian
fit.} \label{fignong}
\end{figure}

As a last characterization of the microscopic dynamics we study the probability distribution of the displacements
$P(dy(t))$ and the non-gaussian parameter $\alpha_{2,y}=\langle dy ^4(t) \rangle/3\langle dy^2(t) \rangle^2 -1$.
The latter characterizes broad, non-gaussian, tails to $P(dy(t))$, reflecting cage rearranging motions such as in
Fig.~1(e). Figure~\ref{fignong}(a) shows $\alpha_{2,y}(t)$ for various $\dot{\gamma}$. It exhibits a peak for
$t\equiv\tau_2$ corresponding to the crossover from caged to diffusive behavior in the MSD (inset,
Fig.~\ref{figversusgammadot}(a)), and vanishes for $t \gtrsim \tau_{\alpha}$. A non-zero $\alpha_2$ also suggests
cooperative motion, consistent with the heterogeneous trajectories for $t \lesssim \tau_\alpha$ in
Fig.~\ref{fig_tra}~(b)-(d). The peak time follows $\tau_2 \propto \dot{\gamma}^{0.65}$,
Fig.~\ref{figversusgammadot}(a), somewhat different from the $\tau_{\alpha}$ scaling. More interestingly, the
distributions $P[dy(t=\tau_2(\dot{\gamma}))]$ show a near collapse for different rates, (Fig.~\ref{fignong}(b)),
despite a slight decrease of $\alpha_2(\tau_2)$ with $\dot{\gamma}$. In quiescent systems at $\phi < \phi_g$, such
(near) collapse of $P[dy(\tau_2(\phi))]$ at different $\phi$ is {\it not} expected since there $\alpha_2(\tau_2)$
grows strongly with $\phi$ while the MSD at $\tau_2$ decreases rapidly \cite{Weeks_Kegel}.

\begin{figure}

\scalebox{0.45}{\includegraphics{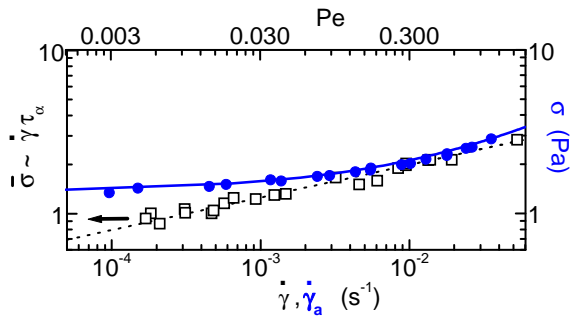}}

\caption{($\square$) Local `flow curve' $\bar{\sigma}=G_0 \dot{\gamma} \tau_{\alpha}$, with $G_0=8.5$~Pa, vs.
$\dot{\gamma}$ (dashed line: $\bar{\sigma} \propto \dot{\gamma}^{0.2}$) compared with the macroscopic flow curve
$\sigma(\dot{\gamma}_{a})$ measured in cone-plate geometry (\Blue{$\bullet$}). Full line: fit to the
Herschel-Bulkley model $\sigma = 1.36 \;\mbox{Pa} + A \dot{\gamma}_{a}^{0.56}$. }
      \label{figflowcurve}
\end{figure}

We now return to the $\dot{\gamma}$ dependence of $\tau_{\alpha}$. There is currently no firm theoretical basis
for relating $\tau_{\alpha}$ to flow properties. Nevertheless, $\tau_\alpha$ is often taken (with some
simulational evidence \cite{VarnikPRB06}) as proportional to viscosity \cite{FuchsCatesPRL02_nonlinearheo}, giving
an effective stress $\bar{\sigma} = G_0 \tau_\alpha \dot{\gamma}$ with $G_0$ a modulus. The resulting
`microscopic' flow curve shows $\bar{\sigma} \propto \dot{\gamma}^{0.2}$, Fig. \ref{figflowcurve}. Recent theories
\cite{SollichCatesPRL97_softglassyrheology, FuchsCatesPRL02_nonlinearheo} have argued for the existence of a
dynamic yield stress at $\dot{\gamma} \rightarrow 0^{+}$ in uniform shear. However, our results show no sign of a
plateau in $\bar{\sigma}$ for reduced rates down to $\mbox{Pe} \simeq 0.005$.

Figure \ref{figflowcurve} shows the experimental {\it global} flow curve measured with a stress controlled
rheometer (AR2000, TA Instruments) in cone-plate geometry (diameter $40$~mm, angle $1^{\circ}$, both surfaces
coated with particles). The stress $\sigma$ is related to the average shear rate $\dot{\gamma}_{a}$ by
$\sigma(\dot{\gamma}_{a})=\sigma_Y^{(D)}+ A \dot{\gamma}_{a}^{~n}$ with a dynamic yield stress
$\sigma_Y^{(D)}=1.36$~Pa and $n=0.56$, similar to previous HS measurements
\cite{PetekidisPuseyJPhCondMat04_shear}.

To compare with the microscopic behavior $\bar{\sigma} = G_0 \tau_\alpha \dot{\gamma}$, in Fig.\ref{figflowcurve}
we have chosen $G_0$ to optimize agreement between $\bar{\sigma}$  and $\sigma(\dot{\gamma}_{a})$ at high
$\dot{\gamma}$. Clearly, the microscopic and macroscopic data disagree. Some discrepancy may be due to the fact
that, for $\mbox{Pe} \gtrsim 1$, hydrodynamic effects render the relation $\bar{\sigma} \propto \tau_{\alpha}
\dot{\gamma}$ less valid. More importantly, discrepancy could arise from the presence of shear localization, e.g.
due to the existence of a static yield stress \cite{Berthier,VarnikBarratPRL03_LJshearlocali}. We have already
mentioned that in our parallel-plate shear cell, a global shear rate $\dot{\gamma}_{a}$ typically corresponds to a
jammed region ($\dot{\gamma} \simeq 0$) coexisting with a flowing region with $\dot{\gamma} > \dot{\gamma}_{a}$.
Preliminary flow imaging {\it inside} our rheometer shows that shear localization also occurs in the cone-plate
geometry, and sets in for $\dot{\gamma}_{a} \leq 10^{-2}$~s$^{-1}$ \cite{fn_microvsmacro}. In LJ simulations
\cite{VarnikBarratPRL03_LJshearlocali}, differences between $\dot{\gamma}$ and $\dot{\gamma}_{a}$ could indeed
explain small deviations between local and global rheology. But the rather larger deviations in
Fig.~\ref{figflowcurve} cannot be explained by this argument. Possibly, the relation $\bar{\sigma} = G_0
\tau_\alpha \dot{\gamma}$ is an oversimplification \cite{footnote} and instead we may need to invoke analogies
with `force chain' dominated systems to make progress. Indeed, our {\it global} shear profile $\dot{\gamma}(z)$,
which exhibits a smooth rather than a step-like decay of $\dot{\gamma}(z)$ to zero (data not shown), has
similarities with velocity profiles in granular matter \cite{Granular}.

Concluding, we have studied the 3D particle dynamics in a HS colloidal glass under steady shear by fast confocal
imaging. Shear occurs in `fluidized' bands where colloids show nearly isotropic `cage breaking' and exponential
relaxation, in contrast to the stretched-exponential dynamics in dense colloidal fluids. The relaxation rate
scales as a power of the {\it local} shear rate: $\tau_\alpha^{-1} \propto \dot{\gamma}^{0.8}$. The `na\"{\i}ve'
microscopic flow curve deduced from this result differs from the global, Herschel-Bulkley, rheology. These and
other recent results \cite{confocal} show the potential of fast 3D imaging to address fundamental questions in
non-equilibrium physics.

We thank M.~E.~Cates, M.~Fuchs, L. Isa, A. Morozov, P.N. Pusey, K.S. Schweizer and F. Varnik for discussions, M.
Jenkins for providing his coordinate refinement routine, and EPSRC GR/S10377 and EP/D067650 (UK) and NSF
DMR-0603055 (US) for funding.



\begin{references}

\bibitem{molglass}M.D. Ediger {\it et al.}, J. Phys. Chem.
{\bf 100}, 13200 (1996).

\bibitem{metglass}M. Heggen {\it et al.}, J. Appl. Phys. {\bf 97}, 033506 (2005); H. Kato {\it et al.}, Appl. Phys. Lett. {\bf
73}, 3665 (1998).

\bibitem{rheoLarsonBook}R.G. Larson, {\it The Structure and Rheology of Complex Fluids}
(Oxford University Press, New York, 1999).

\bibitem{SollichCatesPRL97_softglassyrheology}P. Sollich {\it et al.}, Phys. Rev. Lett. {\bf 78}, 2020 (1997).

\bibitem{Berthier}L. Berthier, J. Phys. Cond. Mat. {\bf 15}, S933 (2003).

\bibitem{FuchsCatesPRL02_nonlinearheo}M. Fuchs and M.E. Cates, Phys. Rev. Lett. {\bf 89}, 248304
(2002); Far. Disc. {\bf 123}, 267 (2003).


\bibitem{Falk}M.L. Falk and J.S. Langer, Phys. Rev. E {\bf 57}, 7192 (1998).

\bibitem{YamamotoOnukiPRE98PRL98}R. Yamamoto and A. Onuki, Phys. Rev. E {\bf 58}, 3515
(1998); Phys. Rev. Lett. {\bf 81}, 4915 (1998).

\bibitem{BerthierJChemPhys02}L. Berthier and J.L. Barrat, J. Chem. Phys. {\bf 116}, 6228 (2002).

\bibitem{Miyazaki04} K. Miyazaki {\it et al.}, Phys. Rev. E {\bf 70}, 011501 (2004).

\bibitem{Petekidis}G. Petekidis {\it et al.}, Phys. Rev. E {\bf 66}, 051402 (2002).

\bibitem{Hebraud}P. Hebraud {\it et al.}, Phys. Rev. Lett. {\bf 78}, 4657 (1997).

\bibitem{rheoimage} J. Lauridsen {\it et al.}, Phys. Rev. Lett. {\bf 93},
018303 (2004). P. Varadan and M.J. Solomon, J. Rheol. {\bf 47}, 943 (2003); D. Derks {\it et al.}, J. Phys. Cond.
Mat. {\bf 16}, 3917 (2004); I. Cohen {\it et al.}, Phys. Rev. Lett. {\bf 93}, 046001 (2004).

\bibitem{BonnPRL02_laponiteagingrejuvenationrheo}D. Bonn {\it et al.}, Phys. Rev. Lett. {\bf 89}, 015701 (2002).

\bibitem{CoussotBecubanding}P. Coussot {\it et al.}, Phys. Rev. Lett. {\bf 88} 218301 (2002);
L. Becu {\it et al.}, Phys. Rev. lett. {\bf 96}, 138302 (2006).

\bibitem{VarnikBarratPRL03_LJshearlocali}F. Varnik {\it et al.}, Phys. Rev. Lett. {\bf 90}, 095702  (2003).

\bibitem{fn_polydisp}Deduced from slow quiescent crystallization kinetics.

\bibitem{YethirajNature03_CHBmodelsystem}A. Yethiraj, A. van Blaaderen, Nature {\bf 421}, 513 (2003).

\bibitem{fn_expversusHS} $\sqrt{\langle dy^2 \rangle}_{\phi_g} \simeq 0.13a$; for pure HSs
\cite{MegenPRE98_tracersinglass} $\sqrt{\langle dy^2 \rangle}_{\phi_g} \simeq 0.18a$.

\bibitem{CrockerGrierJColIntSc96_tracking}J.C. Crocker and D.G. Grier, J. Col. Int. Sc. {\bf 179}, 298 (1996);
M. Jenkins (private communication). At our frame rate, shear-induced distortions in the $x, z$ plane are
unimportant for $\dot{\gamma} \lesssim 0.05$~s$^{-1}$.

\bibitem{fn_localorder}P.R. ten Wolde {\it et al.}, J. Chem. Phys {\bf 104}, 9932 (1996); The average number of
crystalline bonds per particle was $\langle N_x \rangle \sim 2.5$ and constant in time, with only small
(fluctuating) clusters of crystalline particles (with $N_x \geq 8$ ).

\bibitem{HSshearorder}B.J. Ackerson and P.N. Pusey, Phys. Rev. Lett. {\bf 61}, 1033 (1988);
M.D. Haw {\it et al.}, Phys. Rev. E {\bf 57}, 6859 (1998). We observed (partial) crystallization for
$\dot{\gamma} \gtrsim 0.1$~s$^{-1}$.

\bibitem{Weeks_Kegel}E.R. Weeks {\it et al.}, Science {\bf 287}, 627 (2000); W.K. Kegel and A. van Blaaderen, Science
{\bf 287}, 290 (2000).

\bibitem{CourtlandWeeksJPCM03_agingglassconfocal}R.E. Courtland and E.R. Weeks, J. Phys. Cond. Mat. {\bf 15} S359 (2003).

\bibitem{MegenPRE98_tracersinglass}W. van Megen {\it et al.}, Phys. Rev. E {\bf 58}, 6073 (1998).

\bibitem{fn_tauaphi}Remarkably, Fig. \ref{figversusgammadot}(a) includes two points at lower $\phi$,
$\phi[10^3\dot{\gamma}=0.47$~s$^{-1}] \simeq 0.61$ and $\phi[10^3 \dot{\gamma}=1.48$~s$^{-1}] \simeq 0.60$.

\bibitem{Horner96}H. Horner, Z. Phys. B {\bf 100}, 243 (1996).

\bibitem{KobelevSchweizer_shear}V. Kobelev and K.S. Schweizer, Phys. Rev E. {\bf 71}, 021401
(2005); data re-plotted from their Figs. $9$ and $13$.

\bibitem{VarnikPRB06}F. Varnik, O. Henrich, Phys. Rev. B. {\bf 73}, 174209 (2006).

\bibitem{fn_tauj}Taking $\tau_{\alpha,j}$ evaluated from
$F_s(Q_m \parallel j,t=\tau_{\alpha,j})=1/e$, yields $D_j \tau_{\alpha,j} \simeq 0.11$ independent of $j$.

\bibitem{FossBradyJFM99_sheardiffusion}D.R. Foss and J.F. Brady, J. Fl. Mech. {\bf 401}, 243 (1999).

\bibitem{QiuChaikinPRL88_diffusioninshear}X. Qiu {\it et al.},  Phys. Rev.
Lett. {\bf 61}, 2554 (1988).

\bibitem{BreedveldJCP02_nonbrowndiffusion}V. Breedveld {\it et al.}, J. Chem Phys. {\bf
116}, 10529 (2002).

\bibitem{PetekidisPuseyJPhCondMat04_shear}G. Petekidis {\it et al.}, J. Phys. Cond. Mat. {\bf 16}, S3955 (2004).

\bibitem{fn_microvsmacro}Thus the correspondence between our {\it bulk} rheology
and the predictions in \cite{SollichCatesPRL97_softglassyrheology,FuchsCatesPRL02_nonlinearheo} for {\it uniform}
shear is puzzling.

\bibitem{footnote}An explanation of the difference between $\sigma(\dot{\gamma}_a)$ and $\bar{\sigma}=G_0 \tau_{\alpha} \dot{\gamma}$
in terms of $G_0=G_0[\phi(\dot{\gamma})]$ along with $\phi>\phi_0$ (the average volume fraction) in the sheared
region faces the difficulty that in the jammed regions we would have $\phi < \phi_0$. Further, shear-induced size
segregation is unlikely due to the small measured values of shear induced migration.

\bibitem{Granular}W. Losert {\it et al.}, Phys. Rev. Lett. {\bf 85}, 1428 (2000);
E. Aharonov, D. Sparks, Phys. Rev. E {\bf 65}, 051302 (2002).

\bibitem{confocal} L. Isa {\it et al.}, Phys. Rev. Lett. {\bf 98}, 198305 (2007);
I. Cohen {\it et al.}, Phys. Rev Lett. {\bf 97}, 215502 (2006).

\end{references}
\end{document}